# Statistical Properties of Autonomous Flows in 2D Active Nematics

Linnea M. Lemma[1,3], Stephen J. DeCamp[1], Zhihong You[2], Luca Giomi[2] and Zvonimir Dogic[1,3]

[1] Department of Physics, Brandeis University, Waltham MA 02454
[2] Instituut-Lorentz, Universiteit Leiden, P.O. Box 9506, 2300 RA Leiden, Netherlands
[3] University of California at Santa Barbara, Santa Barbara, CA 93111

**Abstract:** We study the dynamics of a tunable 2D active nematic liquid crystal composed of microtubules and kinesin motors confined to an oil-water interface. Kinesin motors continuously inject mechanical energy into the system through ATP hydrolysis, powering the relative microscopic sliding of adjacent microtubules, which in turn generates macroscale autonomous flows and chaotic dynamics. We use particle image velocimetry to quantify two-dimensional flows of active nematics and extract their statistical properties. In agreement with the hydrodynamic theory, we find that the vortex areas comprising the chaotic flows are exponentially distributed, which allows us to extract the characteristic system length scale. We probe the dependence of this length scale on the ATP concentration, which is the experimental knob that tunes the magnitude of the active stress. Our data suggest a possible mapping between the ATP concentration and the active stress that is based on the Michaelis-Menten kinetics that governs motion of individual kinesin motors.

**Introduction:** Active materials exhibit complex dynamical behaviors that are generated through the continuous motion of the microscopic constituents[1-5]. Similar to their equilibrium counterparts, active matter systems can be classified by the structural and dynamical symmetries of the elemental building blocks[6, 7]. One class of active systems is composed of anisotropic motile units that form orientationally ordered liquid crystalline phases. As in an equilibrium nematic, the molecules in an active nematic align along a common local axis called the nematic director. While conventional nematics attain an equilibrium state by assuming a uniform defect-free alignment, active nematics are inherently unstable and exhibit chaotic autonomous flows. Such dynamics results from the instability of uniformly aligned extensile active nematics that drives the formation of pairs of oppositely charged topological defects[8]. The asymmetric positively charged +½ defects acquire motility and stream throughout the sample before annihilating with their counterparts of the opposite charge[9-13]. In a steady state the rate of defect creation and annihilation are balanced. Dynamics of active nematics have been observed in diverse experimental systems ranging from shaken granular rods, to motile cells, to reconstituted cytoskeletal components[14-17]. However, a quantitative comparison of theoretical models to experimental results remains a significant challenge.

We analyze the self-generated dynamics of two-dimensional active nematics comprised of microtubule (MT) filaments and molecular motor kinesin[18-22], and which is fueled by the ATP hydrolysis. In particular, we quantify large scale dynamics and use topological analysis to identify vortices. Analysis confirms the predicted exponential distribution of the vortex areas, allowing us to extract the active length scale, $l_a$. We vary the ATP concentration and elucidate how this parameter controls the active length scale. Our results suggest a scaling relationship that relates the ATP concentration to the magnitude of the active stresses and relies on the results from simulations and detailed knowledge about the stepping kinetics of kinesin molecular motors extracted from single molecule experiments.

Active nematics are characterized by an inherent length scale, $l_a$, expressing the distance at which the restoring torques originating from the orientational elasticity of the nematic phase balance the hydrodynamic torques fueled by the activity. This length scale can be expressed as the ratio between the Frank elastic constant, $K$, of the nematic fluid, which sets the magnitude of the restoring torques, and the active stress, $\alpha$, sourcing the hydrodynamic flows. It follows that $l_a = \sqrt{K/|\alpha|}$, where the absolute value accounts for the fact that $\alpha$ is positive for contractile systems or negative for extensile systems[23]. The comparison between the active length scale $l_a$ and the confinement length scale, $L$, defines whether the active nematic system forms a stationary state ($l_a \gg L$), a state characterized by spontaneous distortion and laminar flow ($l_a \approx L$), or a chaotic state sometimes referred to as active turbulence ($l_a \ll L$)[24]. The onset of chaos is generally anticipated by oscillatory phenomena, depending upon the system geometry and specific material properties. For fully developed active turbulence, the hydrodynamic theory predicts that the flow forms an ensemble of vortices, whose area follows an exponential probability distribution:

$$n(a) = \frac{N}{Z}\exp(-a/a^*), \qquad a_{min} \leq a < \infty, \tag{1}$$

where $dN = da\, n(a)$ is the total number of vortices whose area is between $a$ and $a + da$[23]. $N$ is the total vortex number, $Z$ is a normalization constant, $a^* \sim l_a^2$ is the characteristic vortex area proportional to the square of the active length scale and $a_{min}$ is the minimum area of the active vortex. The average vortex area is proportional to $a^*$, thus the larger the active stress, the smaller are the vortices and the more peaked in the distribution of their area.

**Experimental Methods**: The active nematics studied were comprised of three components: filamentous MTs, biotin labeled kinesin-motors bound into multimotor clusters by tetrameric streptavidin[25], and a depletion agent that induced passive assembly of MT bundles while still allowing for their relative sliding[26-28]. Kinesin clusters simultaneously bound multiple MTs within a bundle, and depending on their relative polarity, generated active stress through extension. Following previous work, we sedimented extensile bundles onto a surfactant stabilized oil-water interface where they assembled into a dense quasi-2D thin nematic film[19]. The ATP-fueled motion of the kinesin motors powered the continuous streaming dynamics of the active nematic films.

Bovine tubulin was purified and labeled according to the previously published protocol[29]. The kinesin motor protein used was the 401-amino acid N-terminal domain from *Drosophila melanogaster* kinesin-1 that is fused to the *E. coli* Biotin Carboxyl Carrier Protein (BCCP) and labeled with a six histidine tag[30]. K401-BCCP-6HIS was expressed in Rosetta pLysS *E. Coli.* in the presence of biotin and purified on a nickel column. For long term storage kinesin was dialyzed against 50 mM imidozole, frozen in a 36% sucrose solution, and stored at -80° C. Motor clusters were created by incubating streptavidin with biotinylated kinesin for 30 minutes on ice. The regeneration system composed of phosphoenol pyruvate monopotassium salt (Beantown Chemicals, #129745) and pyruvate kinase/lactic dehydrogenase (Sigma Aldrich #P0294) was used to maintain a constant ATP concentration over period of hours, even for concentrations as low as 10 μM. The dynamics of active nematics is highly sensitive to the source and purity of phosphoenol pyruvate. Polyethylene glycol (20 kDa) was added as a depletant. Anti-oxidant solution composed of glucose oxidase (0.27 mg/mL), catalase (47 $\mu$g/mL), glucose (4 mg/mL), DTT (66.5 $\mu$M) and trolox (2 mM)--were used to prevent photobleaching. All of the components were suspended in M2B buffer (80 mM PIPES pH 6.8, 1 mM EGTA, 2 mM $MgCl_2$).

In order to track the flow of the active nematic, a sample of 1.33 mg/mL unlabeled MTs was doped with dilute MTs labeled with the Alexa-647 dye. In the final samples there was one labeled MT for every ~15,000 unlabeled ones (**Fig. 1a, Supplementary Movie 1, 2**). Particle Image Velocimetry (PIV) was used on the speckle pattern generated by the sparse fluorescent MTs to obtain active nematic velocity field. Sparse labeling was necessary as the PIV algorithms failed to accurately measure displacements in the active nematics comprised only of fluorescently labeled MTs. This is especially the case for motion along the nematic director due to the axially-symmetric and uniform pattern of the striated MT bundles.

To create a 2D nematic, a flow cell with dimensions of 18 mm length, 3 mm width, and ~50 μm height was made by sandwiching laser cut spacers between a microscope slide and coverglass. The bottom slide was made hydrophobic with commercially available Aquapel. The cover slide was coated with a poly-acrylamide brush to ensure a passive non-sticky hydrophilic surface[31]. The cell was first filled with perfluorinated oil (HFE-7500, 3M, St. Paul) that was saturated with PFPE-PEG-PFPE fluoro-surfactant (RAN Biotechnologies, Beverly, MA) at 1.8% w/v. Subsequently, the active mixture was flowed through the cell while simultaneously wicking out the oil, resulting in a large, flat surfactant-covered oil-water interface onto which the MT bundles adsorbed. The formation of a uniform nematic layer was aided by centrifugation for 10 minutes at 1000 RPM (Sorval Legend Rotor #6434) to help sediment the MTs onto the oil-water interface. Previous work has shown that the structure of the MT active nematics is highly sensitive on the nematic layer thickness, and for thin layers the anisotropic defects create a supra-nematic phase with long range orientational order[19]. The data obtained for this study consisted of active nematics with a retardance of 0.6±0.1 nm. There is a moderate amount of long-range orientational order for these conditions at saturating ATP concentration. The retardance decreases by as much as 20% over the sample lifetime indicating that the nematic layer gets thinner over time, either due to coarsening or losing the MTs to the bulk.

Active nematics were imaged using a conventional fluorescence microscope (Nikon-Ti Eclipse) equipped with an air objective (20× PlanFluor, NA 0.75). The large chip size of our sCMOS camera (Andor-Neo) allowed us to image the active nematic over the area of 0.832 mm × 0.702 mm. The frame rate was tuned for each ATP concentration to insure that the PIV algorithm could track fluorescent MTs and accurately reconstruct the velocity field. About ~10,000 images were acquired for each ATP concentration, with the exception for samples at very low ATP concentrations, where the sharp drop off in the sample speed set a practical limit to acquisition time.

Differences in protein preparation and chemical stocks lead to different dynamics. To ensure reproducibility we polymerized a single large batch of GMPCPP stabilized MTs that were aliquoted and frozen at -80° C. The labeled MTs were polymerized separately from the unlabeled ones. Additionally, a single large batch of the ATP regeneration system, PEG, kinesin motor clusters and antioxidants was made, aliquoted and frozen at -80° C. On the day of the experiments, the components of the active mixture were mixed according to the following protocol: the MTs and premixture were rapidly thawed (at 0 minutes); the ATP was added at the desired concentration—between 10 $\mu$M and 500 $\mu$M (at 2 minutes); the unlabeled MTs were added (at 5 minutes); the labeled MTs were diluted (at 6 minutes); the dilute labeled MTs were added (at 7 minutes); the sample was flowed into the chamber (at 9 minutes); the chamber was sealed and put into the centrifuge for sedimentation (at 12 minutes). Preliminary tests indicated that the timing differences in the sample preparation protocol could significantly alter the system dynamics. There is latitude in choosing the exact timing in this protocol. However, once chosen it was followed consistently to within 1 minute, producing quantitatively reproducible results.

**Detecting vortices:** The velocity field of active nematics was obtained using a modified version of the MATLAB plugin PIVLab (**Fig. 1b**). To identify vortices and measure their areas, we followed the previously published method[23]. Briefly, we extracted a 2D Okubo-Weiss (OW) field, $Q$, from the measured flow velocity (**Fig. 1c**). $Q$ is defined as:

$$Q(x,y) = -\det[\nabla \boldsymbol{v}(x,y)] \qquad (2)$$

where $\boldsymbol{v}$ is the flow velocity. $Q$ is related to the Lyapunov exponent of tracer particles advected by the flow. Negative $Q$ values indicate that two fluid elements, initially close together, will remain so, while positive $Q$ values imply that the fluid elements diverge from each other with time. Since streamlines around a vortex remain parallel to each other a simply connected region where $Q$ is less than zero is indicative of vortices. We emphasize that not all simply connected regions of the OW field are vortices. To classify a region as a vortex, the flow field associated with the OW field has to contain a singularity. We used a previously developed algorithm to identify all of the singularities in the experimentally measured velocity field[19]. For each singularity in the flow velocity field, the vortex size was determined by the area of the associated simply connected region of the OW field. Measured in this way, the vortex area distributions do not depend on sampling speed or field of view.

The vortex detection algorithm is sensitive to the noise in the experimentally measured velocity flow field. Specifically, the distribution of vortex areas is dependent on the size of the grid onto which the PIV data is interpolated, and from which the Okubo-Weiss field is calculated. On the one hand, if the PIV grid spacing is small, noise in the experimentally measured flow field results in many fragmented small regions with $Q<0$. In this limit, the vortex finding algorithm identifies fictitious vortices, which increase the apparent probability of finding small vortices and skews the measured distribution. On the other hand, large grid spacing coarse grains over experimentally relevant smaller vortices. In this limit, the statistical significance of large-area vortices is over-counted broadening the distribution. To determine the optimal choice of input parameters, we systematically changed the PIV grid size for each ATP concentration (**Fig. S1**). For grid sizes above a critical value we observed appearance of a peak in the vortex area distribution. Similar peak was also observed from calculated flow fields that do not contain experimental noise[23]. Physically, this peak represents the minimum area of a vortex created by the active stresses. Smaller vortices may occur due to shear forces between active vortices. Therefore, for each experimental condition we chose the smallest grid size at which the peak at $a_{min}$ appears. This method yields a good agreement between the visual examination of the flow fields and the corresponding locations of the algorithmically detected vortices. For the lowest ATP concentration (10 μM), peak in the distribution was not observed for any grid spacing. This is likely due to the low statistics which result from vortices spanning the field of view. Therefore, the fictitious vortices due to experimental noise make up a larger portion of the detected vortices. Thus, we are likely underestimating the active length scale at 10 μM ATP concentration.

**Experimental results:** The kinesin speed is determined by the ATP concentration. Specifically, single molecule studies have shown that at low ATP concentrations, kinesin speed increases linearly with increasing ATP concentration, and that above a certain ATP concentration the kinesin speed saturates[32]. These microscopic considerations suggest that ATP concentration could also affect the large-scale structure and dynamics of active nematics. Indeed, in exploratory experiments we found that defect density at low ATP concentrations was significantly lower when compared to samples prepared at high ATP concentrations (**Fig. 2**). These observations suggest that changing the ATP concentration can be used to tune the active stress, which are related to the active nematic length scale by the scaling relationship: $l_a = \sqrt{K/|\alpha|}$. Motivated by these consideration we measured the active length scale and its dependence on the ATP concentration.

We measured the active nematic flow fields at a series of ATP concentrations. Using the above described procedures, we extracted the vortex area probability distribution $n(a)$ (**Fig. 3**). Vortex sizes exhibited an exponential distribution above a critical vortex size, $a_{min}$, for most of ATP concentrations studied. As mentioned previously anomalous distributions at lowest ATP concentrations are because the average vortex size approaches the experimental field of view. We also observed that increasing the ATP concentration leads to the significantly narrower vortex size distributions. Additionally we calculated the mean vorticity of vortices as a function of vortex area, $<\omega_v(a)> = <\nabla \times v(a)>$, where $<>$ denotes averaging over both space and time (inset of **Fig. 3**). The mean vorticity increases with ATP concentration. For all ATP concentrations studied we observed that mean vorticity of different sized vortices remains constant, in agreement with theoretical predictions[23]. The multi-scale structure of turbulence has other consequences on the statistical properties of the active nematic flows. As in inertial turbulence, we found that the measured velocities of the active nematics followed a Gaussian distribution (**Fig. 4a**), which is in agreement with numerical simulations[23]. In comparison the distribution of the measured vorticities exhibited distinct non-Gaussian tails (**Fig. 4b**), which is also found theoretically[23].

The measured velocity fields also yield the equal-time velocity-velocity correlation function functions, $C_{vv}(r) = \langle \boldsymbol{v}(0) \cdot \boldsymbol{v}(\boldsymbol{r}) \rangle / \langle |\boldsymbol{v}^2(0)| \rangle$, and the equal-time vorticity-vorticity correlation function, $C_{\omega\omega}(r) = \langle \omega(0)\omega(\boldsymbol{r}) \rangle / \langle |\omega^2(0)| \rangle$ (**Fig. 5a, 5b**). Care has to be taken when interpreting the calculated vorticity correlation functions. The inherent experimental noise is amplified when taking derivatives to calculate the vorticity correlation functions. This random noise introduces a sharp drop off at small separations of the correlation function so that the normalized correlation function no longer smoothly extrapolated to unity. These effects are especially pronounced at low ATP concentrations, where the average velocities are smaller and noise becomes more pronounced. We corrected for this effect by keeping only the data over the range of values where the correlation function is smooth and shifting it in the y-direction to ensure that it interpolates smoothly to unity at zero spatial separation.

Vortex size distributions, velocity and vorticity correlation functions provide three independent methods of extracting the active length scale, $l_a$, and its dependence on the ATP concentration. As previously discussed the characteristic length of the exponential distribution of vortex areas provides a direct measurement of the active length scale. Alternatively, the length over which both the velocity-velocity and vorticity-vorticity correlation functions decayed to half their maximum provide two other independent methods of extracting the active length scales, $l_a$[33]. We found that the active length scales extracted from these analyses scale similarly (**Fig. 5c**). Specifically, at low ATP concentrations the measured active length scale decreased with increasing ATP, and plateaued above a critical ATP concentration of around ~250 μM. Numerical simulations also find that these length scales collapse when rescaled by a constant[33].

**Dependence of active stresses on ATP concentration:** The main parameter that controls the dynamics of active nematics is the magnitude of the active stress, $\alpha$, which can be either extensile or contractile. Hydrodynamic simulations revealed that increasing $|\alpha|$ in an extensile system leads to narrower distribution of vortex sizes and decreases the mean vortex sizes, a trend similar to what is observed experimentally with increasing ATP concentrations[23]. However, relating active stresses generated by kinesin motors to ATP concentrations is challenging. Basic thermodynamic considerations suggest that the magnitude of the active stress scales as the logarithm of ATP concentrations: $\alpha \sim \log[\text{ATP}]$, an assumption used previously[10, 20]. This assumption is rooted in the consideration that the speed at which a kinesin molecule moves along a microtubule is proportional the rate of ATP hydrolysis, which, in turn, is proportional to the difference of the ATP chemical potential before and after hydrolysis, i.e. $\alpha \sim \Delta\mu$. Assuming thermodynamic equilibrium and differentiating the free-energy $F = U + k_B T \sum_{i \in \{\text{ATP}, \text{ADP}\}} N_i \log(N_i/N)$, with $N = N_{\text{ATP}} + N_{\text{ADP}}$ the total number of ATP and ADP molecules, such that $[i] = N_i/N$, yields: $\Delta\mu = (\partial F/\partial N_{\text{ATP}})_{T,N} - (\partial F/\partial N_{\text{ADP}})_{T,N} = k_B T (\log N_{\text{ATP}} - \log N_{\text{ADP}}) \approx k_B T \log N_{\text{ATP}}$, since it is assumed $N_{\text{ATP}} \gg N_{\text{ADP}}$ in our system. This argument relates the active stress directly to the kinetics of the ATP hydrolysis, but ignores how efficiently kinesin converts chemical energy into mechanical work.

We propose a different approach based on a combination of numerical results, expressing the relation between the active stress and the extension rate of the microtubule bundles, and empirical evidence, concerning the kinesin duty cycle. In the experimental realization of the active nematics, internal stresses are generated by kinesin clusters which slide MT bundles. We postulate that the active stress scales as a power of the filament extension rate $v$: i.e. $\alpha \sim v^\beta$. We expect a power law scaling to be valid for different active systems, with the exponent $\beta$ likely being sensitive to microscopic details. This assumption is supported by the results from computer simulation model that is described in the subsequent section. Furthermore, we expect the extension rate $v$ to be proportional to the velocity $V$ of the kinesin motors: i.e. $v \sim V$. The latter, in turn, is known to depend on the ATP concentration by the Michaelis-Menten relation:

$$\frac{V}{V_{max}} = \frac{[\text{ATP}]}{K_m + [\text{ATP}]}, \qquad (3)$$

where $V_{max}$ is the maximal speed attained at ATP saturation, and $K_m$ is the ATP concentration at which the kinesin speed is $V_{\text{max}}/2$[32]. Optical tweezer based experiments reveal that both $K_m$ and $V_{max}$ depend on the magnitude of the force that is applied in the direction opposite of the kinesin movement[34]. Combining these considerations yields:

$$\alpha \sim \left(\frac{[\text{ATP}]}{K_m + [\text{ATP}]}\right)^\beta. \qquad (4)$$

This expression fits reasonably well to the experimental estimate of how the active stress scales with the ATP concentration (**Fig. 6**). The values obtained from the fit are: $K_m$=252$\pm$395µM and $\beta$=0.97$\pm$0.57. The value of the $K_m$ should be taken with caution as this parameter is particularly flexible in this model because of the form of the fit equation. Single molecule experiments showed that $K_m$ depends on the force applied on the motor in the direction opposite of its movement[34]. For vanishing loads $K_m$~90 µM, while just below the stall force $K_m$~320 µM.

The proposed relationship has well-defined limiting behaviors depending on the whether the kinesin clusters operate far or very close to ATP saturation. For $[ATP] \ll K_m$ the kinesin speed is linear in $[ATP]$ and $\alpha \sim [ATP]^\beta$. Close to saturation, $[ATP] \gg K_m$ and:

$$\alpha \sim 1 + \beta \log\left(\frac{[ATP]}{K_m + [ATP]}\right), \tag{5}$$

where we used the expansion $x^\beta \approx 1 + \beta \log x$, for $x \approx 1$. Although both regimes differ from the previously used assumption $\alpha \sim \log[ATP]$[10, 20], they might be difficult to distinguish if the explored range of the ATP concentrations is not sufficiently broad compared to $K_m$.

**Computer simulations:** To better understand how activity scales with ATP concentration we performed coarse-grained molecular dynamics simulation of extensile rods. In the simulation, each bundle was treated as a two-dimensional sphero-cylinder with fixed diameter $d_0$ and time-dependent length $l$ (included the caps), extending on the plane with periodic boundary conditions (**Fig. 7a**). The position, $r_i$, and the orientation, $p_i = (\cos\theta_i, \sin\theta_i)$, of the $i$−th bundle ($i = 1, 2 \ldots$), were governed by the over-damped Newton equations for a rigid body, namely:

$$\frac{d\boldsymbol{r}_i}{dt} = \frac{1}{\zeta l_i}\sum_j \boldsymbol{F}_{ij}, \qquad \frac{d\theta_i}{dt} = \frac{12}{\zeta l_i^3}\sum_j (\boldsymbol{r}_{ij} \times \boldsymbol{F}_{ij}) \cdot \hat{\boldsymbol{z}}, \tag{6}$$

where the summation runs over all the bundles in contact with the $i$−th. The points of contact have positions $\boldsymbol{r}_{ij}$ with respect to the center of mass of the $i$−th bundle and apply Hertzian forces of the form $\boldsymbol{F}_{ij} = E d_0^{1/2} h_{ij}^{2/3} \boldsymbol{N}_{ij}$, where $E$ is an elastic constant, $h_{ij}$ is the overlap distance between the $i$−th and $j$−th bundles and $\boldsymbol{N}_{ij}$ their common normal unit vector. The buffer fluid was not explicitly simulated, but its effect on the dynamics of the bundles is taken into account through the constant $\zeta$, representing the Stokes drag per unit length originating from the solvent. The length $l_i$ increased linearly in time. After reaching a maximal length $l_{\max}$, the bundle was divided in two identical halves and one of them was removed to keep the total particle number constant. In order to avoid synchronization of divisions, the extension rate of each cell, defined as the length increment per unit time, is randomly chosen from an interval $\left[\frac{v}{2}, \frac{3}{2}v\right]$, where $v$ is the average extension rate. This approach, already used to investigate the orientational properties of active nematic defects[35], was not aimed to accurately reproduce the microscopic dynamics of the MT bundles, but rather to provide generic insight into the relation between the active stresses and the extension rate.

The stress tensor experienced by the *i*-th bundle was calculated using the virial expansion, namely;

$$\sigma_i = \frac{1}{2a_i'}\sum_j r_{ij}F_{ij}, \qquad (7)$$

where $a_i' = d_0 l_i/\phi$, with $\phi$ the packing fraction, is the effective area occupied by the bundle. The stress tensor was decomposed in a longitudinal ($\sigma_\parallel$) and a transverse ($\sigma_\perp$) component with respect to the direction of the bundles[36]. From these, one can calculate the pressure $P$ and the deviatoric stress $\alpha$, namely:

$$\boldsymbol{\sigma} = \sigma_\parallel \boldsymbol{nn} + \sigma_\perp \boldsymbol{n_\perp n_\perp} = -P\boldsymbol{I} + \alpha\left(\boldsymbol{nn} - \frac{1}{2}\boldsymbol{I}\right) \qquad (8)$$

where $\boldsymbol{n}$ is the nematic director, corresponding to the average direction of the bundles, $\boldsymbol{n_\perp}$ is a unit vector perpendicular to $\boldsymbol{n}$, $\boldsymbol{I}$ is the identity tensor in two dimensions, $P = -|\sigma_\parallel + \sigma_\perp|/2$ and $\alpha = (\sigma_\parallel - \sigma_\perp)$. After a short transient, the system reaches a steady state in which all the components of the stress tensor fluctuate about a time-independent mean value (**Fig. 7b**). All the components of the stress tensor increased monotonically with the extension rate $v$ (**Fig. 7c**). In particular, the deviatoric stress, $\alpha$, is found to have a power-law dependence on the extension rate, namely: $\alpha \sim v^\beta$, with $\beta \approx 0.314$. This exponent is likely not universal and, in suspensions of MTs and kinesin, is expected to depend on various microscopic details. Nevertheless, these details affect the dependence of the active stress on the ATP concentration only through pre-factors of secondary importance.

**Discussions and Conclusions:** Using a model system of MT based active nematics we demonstrated that the vortex sizes of the autonomous flows follow an exponential distribution, thus providing experimental evidence for the existence of a single active length scale. We extracted the dependence of the active length scale on the ATP concentration from both the size distribution of the vortex areas, and the related velocity and vorticity correlation functions. These results revealed that the characteristic length scale decreases with increasing ATP concentration, in qualitative agreement with scaling arguments. Intriguingly, previous experiments have also measured the characteristic length scale of three dimensional active isotropic fluids, finding that this length scale is largely independent of the ATP concentration[18, 37], suggesting a fundamental difference between these systems.

Our experiments illustrate two features that make MT-based active nematics a unique system for testing theoretical models of active liquid crystals[18, 19, 38]. First, highly efficient molecular motors power non-equilibrium steady state dynamics that persist for multiple hours or even days, allowing one to image dynamics over extended time, making it feasible to obtain large data sets that are required for extracting quantitative measurement of the vortex size distribution, especially at low ATP concentrations. Second, being assembled from well-characterized biochemical constituents MT based active nematic also allow one to systematically tune the microscopic parameters such as the nematic layer thickness and the ATP concentration that determines the velocity of the motor proteins.

Our work also highlights challenges in quantitatively interpreting the dynamics of MT based active nematics. Specifically, we address the problem of tuning the magnitude of the active stress by changing the ATP concentration. The previously discussed generic thermodynamic argument suggests that active stress should be related to the logarithm of the ATP concentration. However, this argument is complicated by the microscopic realities of the kinesin motors. The logarithm of ATP is the energy available to the system, but the efficiency by which the kinesin motors transfer energy into the active nematic depends on the average load applied on the motors. At large loads close to the stall force (~5pN) kinesin motors have a peak efficiency of ~30%. This efficiency decreases significantly with decreasing load[39]. There are no estimates of the average load on the kinesin motors in an active nematic. Thus, it is not possible to estimate how much of the available energy from the ATP hydrolysis is converted into mechanical work and how much is dissipated away through other pathways. From a different perspective, we also note that the individual motors obey the Michaelis-Menten kinetics. Thus, their speed increases linearly with ATP in the low concentration limit and saturates at high concentrations. It is reasonable to assume that these observations are also true for the bundle extension speed, and our simulations demonstrate that active stresses increase as a power law of the bundle extension speed. These microscopic considerations are fundamentally incompatible with the previously proposed logarithmic scaling.

The dependence of the active length scale $l_a \approx \sqrt{K/|\alpha|}$, implies that the active stress $\alpha$ scales like $1/l_a^2$. Thus, plotting $1/l_a^2$ versus ATP concentration estimates how the magnitude of the active stress scales with the ATP concentration (**Fig. 6a**). Another measure of the active stress comes from the average vorticity of vortices, $<\omega>_v$. The balance of viscous and active stress over the size of a vortex implies $<\omega>_v = \alpha/\eta$ where $\eta$ is the shear viscosity[23], which predicts that the active stress $\alpha$ should also scale like $<\omega>_v$. Plotting the two estimates of the active stress against each other yields a linear relationship, which confirms the consistency of our scaling arguments (**Fig. 6b).**

Some caution is needed when interpreting these scaling relationships. The assumption underlying the above arguments is that $K$ and $\eta$ are independent of the ATP concentration. A possible complication arises because in the absence of ATP, kinesin motors bind MTs in a rigor configuration, thus acting as permanent crosslinkers[40, 41]. This suggests that upon ATP depletion the 2D MT layer becomes elastically stiffer and becomes a cross-linked solid rather than an equilibrium nematic fluid. This is consistent with the observation that after the ATP is consumed the defects in the nematic layer remain permanently frozen. In comparison, for an equilibrium nematic all of the topological defects would annihilate with each other to minimize elastic distortions. It is likely that similar considerations are also relevant at low ATP concentrations, where the motors take very few steps each second. Thus, for the majority of the time they remain in the passive state where they are attached to both MTs, and act as conventional cross-linkers that modify the nematic elasticity. If we revert back to the assumption that activity scales as log [ATP] and recall that $l_a = \sqrt{K/|\alpha|}$, then $K \sim l_a^2 \log [\text{ATP}]$. When plotted in this way we find that *K* systematically increases at low ATP concentration (**Fig. S2**).

The exponential distribution of the vortex sizes has been measured previously in both dense cellular tissues and for a MT based active nematics at a saturating ATP concentration[16, 42]. The latter work estimated that an active length scale is $l_a = 24.2 \pm 0.35\,\mu m$ at 700 µM ATP, which is smaller than any of the length scales measured here. One possible reason for this discrepancy are the differences in the sample preparation: the oil used in our work is $10^{-4}$ as viscous and our chamber construction is different. Additionally, there are differences in the details of the analysis. As discussed previously, the distribution of vortex sizes depends on the grid spacing, and one needs to employ a self-consistent method of choosing the appropriate scale. Measuring vortex size distribution requires one to classify the domains of the OW field according to their topological properties. Only regions that have a net charge can be classified as vortices.

Active nematics are highly dynamic materials whose large-scale structure is determined by a characteristic length scale. Following protocols that were initially developed for computational work, we demonstrated an experimental method that extracts the active nematic length scale. This length scale is controlled by tuning the ATP concentration. We also emphasized the importance and challenges of mapping parameters under experimental control, such as ATP concentration, onto theoretically relevant parameters, such as the active stress. Establishing quantitative relationships between these quantities is an essential stepping stone for quantitatively testing various theoretical models of active nematics.

**Acknowledgments:** This work was supported by Department of Energy, Office of Basic Energy Sciences under Award DE-SC0010432TDD. We also acknowledge use of biomaterials and optical microscopy facilities at Brandeis that are supported by NSF-MRSEC-1420382. LG and ZY are supported by The Netherlands Organization for Scientific Research (NWO/OCW) as part of the Frontiers of Nanoscience program and the Vidi scheme.

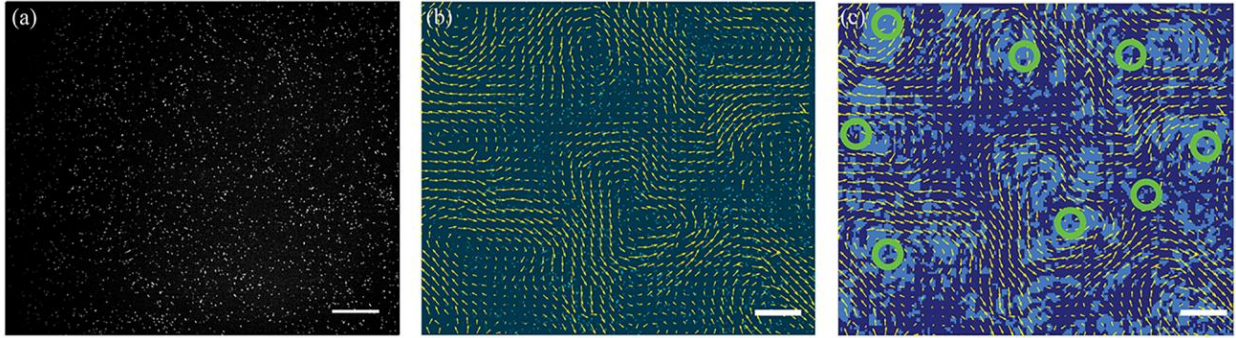

**Figure 1: Vortices in a 2D active nematic flow field.** (a) An active nematic in which one of every ~15,000 MTs is fluorescently labeled. The resulting speckle pattern is suitable for quantifying the active nematic flow field using particle imaging velocimetry (PIV). (b) Velocity field obtained from the PIV analysis overlaid on a raw image of an active nematic containing sparsely labeled microtubules. (c) The Okubo-Weiss field extracted from the velocity field. Light shading specifies areas where $Q<0$, which indicates coherent flows. Dark shading shows regions where $Q>0$, which indicates diverging flows. The green circles indicate regions of coherent flows where there are vortices as defined by velocity field rotations of $2\pi$. The area of the vortex is defined by the sum of the connected areas of $Q<0$ around a vortex center. Scale bars, 100μm.

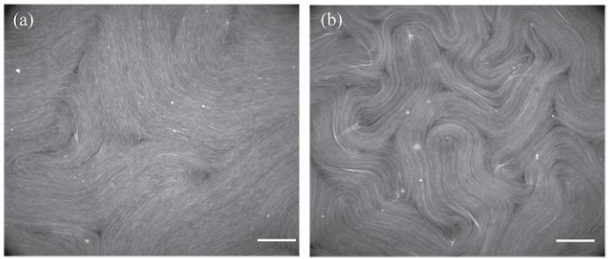

**Figure 2: ATP concentration controls the active nematic length scale.** (a) Image of an active nematic with all the MTs labeled at low ATP concentration (10 μM) where the average defect spacing is large. (b) High ATP concentration lead to smaller active nematic length scale, as evidenced by higher density of defects in the field of view (150 μM). Scale bars, 100 μm.

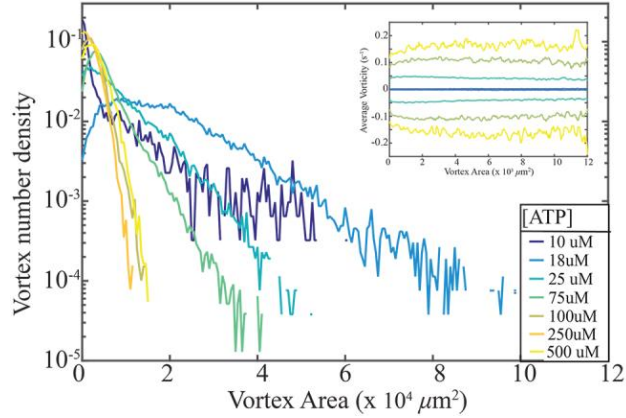

**Figure 3: ATP concentration controls distribution of vortex sizes.** Density of vortices plotted as function of vortex area of 2D active nematic systems plotted for a range of ATP concentrations. The distributions are exponential in a range $a_{min} < a < a_{max}$, where $a_{min}$ is the minimum area of an active vortex. The distribution broadens with decreasing activity (ATP concentration) as predicted by the theory. Inset: Mean vorticity of a vortex as a function of its area plotted for active nematics samples at different ATP concentrations.

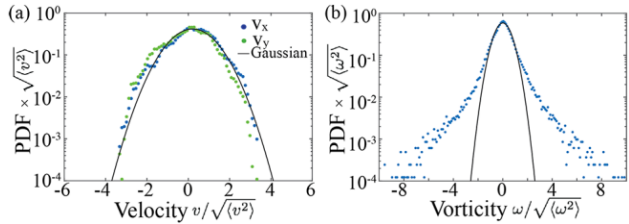

**Figure 4: Velocity and vorticity probability distributions: (a)** Probability distribution function of the velocity components **(b)** Probability distribution function of vorticity. Data is normalized by their corresponding standard deviation. The fit of the data to a Gaussian is shown by the black line. The velocity components PDFs follow a Gaussian distribution, while the vorticity PDF shows deviation from Gaussianity at the tails.

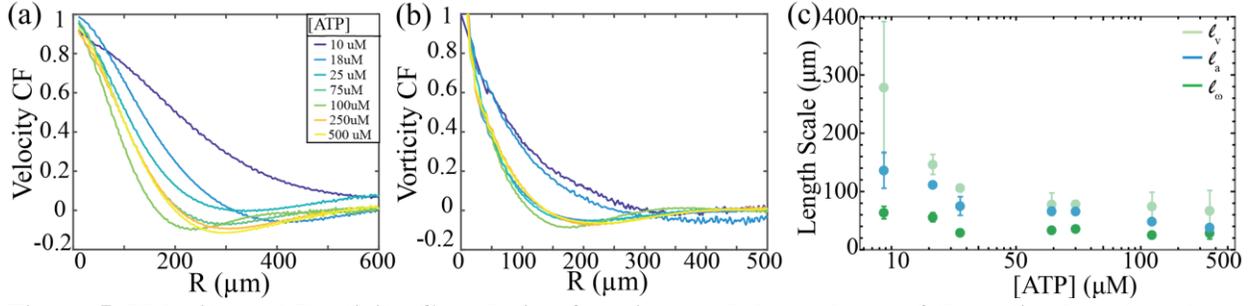

**Figure 5: Velocity and Vorticity Correlation functions and dependence of the active length scale on the ATP concentration.** (a) Velocity-velocity correlation function measured for different ATP concentrations. (b) Vorticity-vorticity correlation function measured for different ATP concentrations. (c) The active length scale as a function of the ATP concentration extracted from the velocity correlation functions (mint), vortex area distributions (blue) and vorticity correlation functions (green). The error bars are the standard deviation of the value of the length scale extracted from multiple samples. All the length scales exhibit the same trend, decreasing with increasing ATP concentration until saturation.

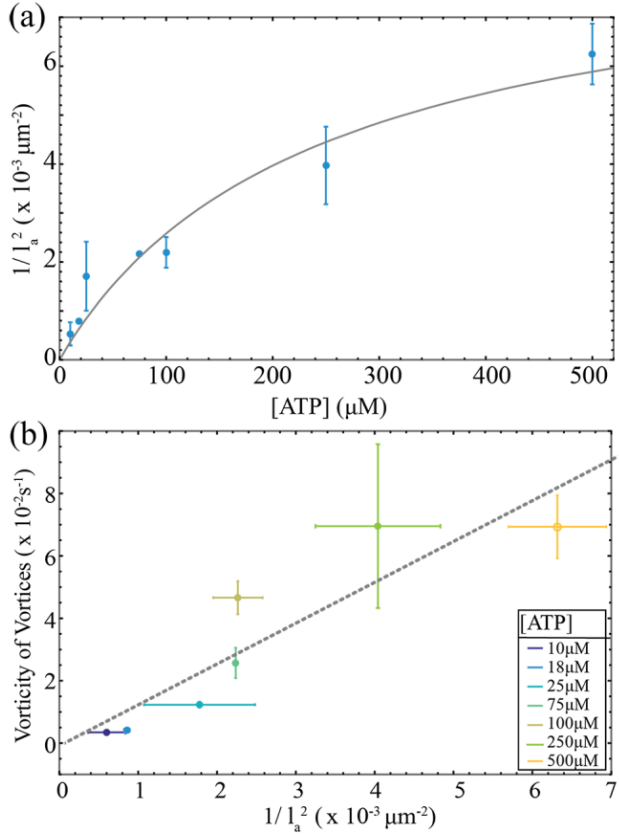

**Figure 6: Dependence of the active stress on the ATP concentration.** (a) The estimate of the the active stress, $1/l_a^2$, as a function of the ATP concentration. The full line indicates the theoretical fit of the active stress which scales as $\alpha \sim v^\beta$, where $v$ is the filament extension velocity that is described by the Michaelis-Menten kinetics (Eq. 5). The experimental fit parameters are $K_m$=252 $\pm$ 394µM and β=0.97$\pm$0.57 where the error is the standard error on the fit parameters. The active length scale, $l_a$, is extracted from the vortex size distributions measured at different ATP concentrations. The error bars are the standard deviation of active length scale from different experiments. (b) Comparison of two different methods of extracting active stresses. One method relies on the estimate of the active stress from vorticity relationship $\alpha \sim \eta <\omega>_v$, while the other uses: $\alpha \sim \frac{K}{l_a^2}$. The colors indicate the ATP concentration at which the two measurements are compared. Gray line is a linear fit of the data.

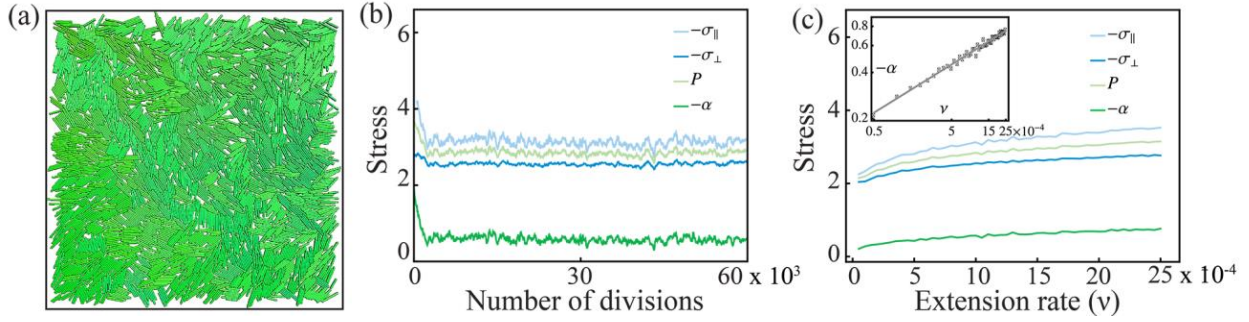

**Figure 7: Dependence of active stresses on extension speed extracted from numerical simulations.** **(a)** Snapshot of the numerical simulations. Microtubule bundles are modeled as spherocylinders whose length $l$ extends linearly in time. Once it reaches the maximum value $l_{\max} = 5d_0$, with $d_0$ the diameter, the bundle is divided in two identical halves and one of them is removed from the system in order to keep particle concentration constant. **(b)** The components of the stress tensor versus time (measured in terms of number of divisions). The four curves represent the longitudinal ($\sigma_\parallel$) and transverse ($\sigma_\perp$) component of the stress tensor, whereas $P = -|\sigma_\parallel + \sigma_\perp|/2$ and $\alpha = (\sigma_\parallel - \sigma_\perp)$ are the pressure and deviatoric stress respectively. Stresses are measured in units of the elastic constant $E$ of the bundles. **(c)** Stress as a function of the bundle extension rate $v$. The latter is expressed in units of $d_0/\tau$ with $\tau = \zeta/E$ the time scale arising from Eq. (3). All the components of the stress increase monotonically with $v$. The deviatoric stress, in particular, exhibits power-law dependence: $\alpha \sim v^\beta$ with $\beta \approx 0.314$ (inset).

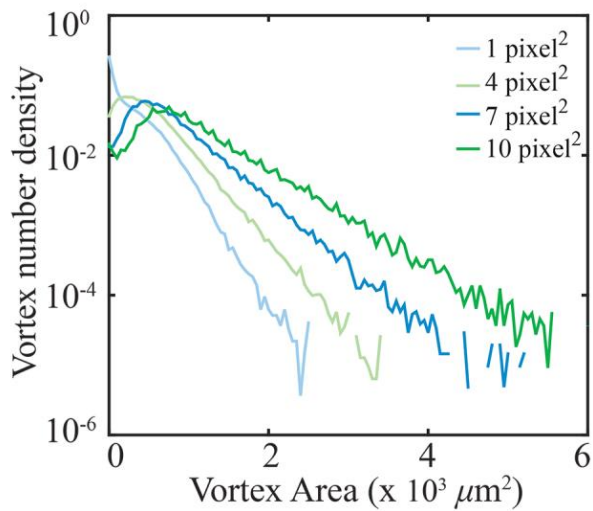

**Figure S1: Distribution's Dependence on Okubo-Weiss Spacing** For a single sample (100μM), the vortex area distributions obtained from different grid spacing. The size of the grid which the PIV and OW field are interpolated on affect the broadness of the distribution and therefore the measured active length scale. To be consistent across samples we took the grid spacing at which a peak at $a_{min}$ was just recovered—in this example 4 pixel$^2$. This corresponds well to tuning the spacing "by-eye" so that the algorithm identifies vortices only where one can see vortices in the field.

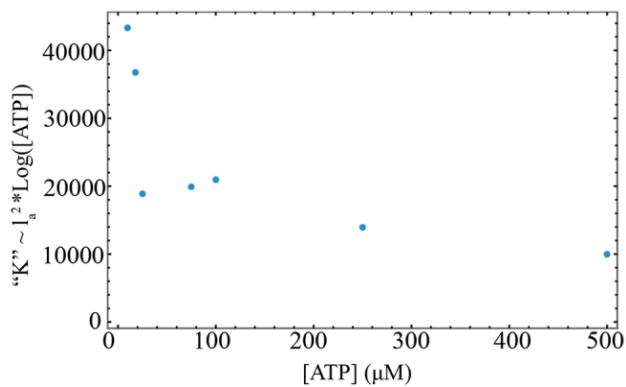

**Figure S2: "K" vs log([ATP])** The dependence of the elastic constant, K on ATP concentration if we assume that $\alpha \sim \log([ATP])$ so that $K \sim l_a^2 \log([ATP])$.

**Supplementary Movie 1:** Raw data of an active nematic at 10 $\mu$M ATP in which ~1:15,000 MTs is fluorescently labeled. The speckle pattern makes images suitable for extracting the velocity field using the particle image velocimetry. Movie is a representative snapshot of a much longer acquisition. Scale bar 100 $\mu$m.

**Supplementary Movie 2:** Raw data of an active nematic at 1000 $\mu$M ATP in which ~1:15,000 MTs is fluorescently labeled. The speckle pattern makes images suitable for extracting the velocity field using the particle image velocimetry. Scale bar 100 $\mu$m.

18